\documentclass[conference]{IEEEtran}
\IEEEoverridecommandlockouts

\usepackage{cite}
\usepackage{amsmath,amssymb,amsfonts}
\usepackage{upgreek}
\usepackage{algorithmic}
\usepackage{graphicx}
\usepackage{textcomp}
\usepackage{xcolor}
\usepackage{booktabs}
\usepackage{array}
\usepackage{multirow}
\usepackage{siunitx}
\usepackage{url}
\usepackage[hidelinks]{hyperref}
\usepackage{threeparttable}
\newcolumntype{C}[1]{>{\centering\arraybackslash}p{#1}}
\newcolumntype{R}[1]{>{\raggedleft\arraybackslash}p{#1}}

\newcommand{\hmin}{\ensuremath{h_{\min}}}
\newcommand{\favg}{\ensuremath{f_{\mathrm{avg}}}}
\newcommand{\Cg}{\ensuremath{C_g}}
\newcommand{\Rg}{\ensuremath{R_g}}
\newcommand{\Rf}{\ensuremath{R_f}}
\newcommand{\degC}{\ensuremath{^{\circ}\mathrm{C}}}

\def\BibTeX{{\rm B\kern-.05em{\sc i\kern-.025em b}\kern-.08em
    T\kern-.1667em\lower.7ex\hbox{E}\kern-.125emX}}

\begin{document}

\title{Gain-Dependent Entropy Characterization of a Zener Avalanche Noise TRNG via Dual NIST Standards: Toward Low-Cost Hardware Entropy for ICS Deployments}

\author{
\IEEEauthorblockN{Rishabh Das}
\IEEEauthorblockA{J. Warren McClure School of Emerging Communication Technologies\\
Scripps College of Communications\\
Ohio University, OH, USA\\
Email: rishabh.das@ohio.edu}

\thanks{The authors gratefully acknowledge the internal funding provided by Scripps College of Communications to support this work.}}
\maketitle

\begin{abstract}
True random number generators based on Zener avalanche noise offer a
low-cost entropy source for industrial control system and IoT edge
devices, yet the relationship between amplifier gain, coupling
capacitance, and entropy quality remains uncharacterized under
current NIST standards. This paper presents a comprehensive entropy
assessment of an op-amp-based Zener avalanche noise generator across
five gain configurations and two coupling capacitor values,
evaluated against the NIST SP~800-22 statistical test suite and the
SP~800-90B entropy source assessment using $1.44\!\times\!10^{9}$ raw bits per configuration. We identify a three-phase autocorrelation
structure, attributed to comparator saturation-recovery dynamics,
consistent across all gain settings. A direct-current substitution experiment, replacing the Zener diode
with a stable voltage source at the same bias point, reduces
switching activity by approximately 97\%, confirming that Zener
avalanche noise, rather than amplifier self-excitation, is the
necessary entropy source. The coupling
capacitor is further shown to govern comparator overdrive and entropy
quality independent of this source mechanism, with the smaller
capacitance improving entropy, throughput, and bias. The combined
results identify an optimal gain configuration achieving a perfect
statistical test score, NIST-compliant min-entropy, and peak entropy
throughput at a bill-of-materials cost under five dollars, offering
practical design guidance for low-cost entropy sources in industrial
and embedded deployments.
\end{abstract}

\begin{IEEEkeywords}
True random number generator, Zener avalanche noise, gain
characterization, NIST SP~800-22, NIST SP~800-90B, entropy source
assessment, von Neumann extractor, industrial control systems,
hardware security.
\end{IEEEkeywords}

\section{Introduction}\label{sec:intro}

True random number generators based on physical noise sources occupy a
central role in hardware security. The random number generators provide the entropy for
cryptographic key generation, nonce production, and challenge--response
authentication. Unlike pseudorandom number
generators, TRNGs derive unpredictability from irreducible physical
phenomena such as thermal fluctuations, shot noise, and quantum
tunneling, and are therefore resistant to algorithmic reconstruction by an adversary who knows the seed state. The practical challenge in TRNG engineering
is not the existence of physical noise but rather the rigorous,
standards-compliant characterization of how much entropy that noise
actually delivers after digitization and conditioning.

Zener avalanche noise has attracted sustained interest as an entropy
source precisely because of its simplicity~\cite{taylor2011, tkacik2002}.
A reverse-biased $p$--$n$ junction operating in the avalanche-breakdown
regime generates broadband shot-noise currents that, when amplified and
thresholded by a comparator, produce a binary switching waveform. The
resulting circuit requires only a Zener diode, an operational amplifier,
and a handful of passive components. 
While oscillator-based and integrated-circuit TRNG topologies have
received extensive characterization~\cite{sunar2007, bucci2003,
baudet2011}, the Zener avalanche-noise family has not: the literature
contains no systematic characterization of how amplifier gain affects
entropy quality in this topology. Published designs typically select a gain empirically and
validate the output with a subset of NIST~SP~800-22 tests~\cite{nist80022},
without examining the gain--entropy relationship or applying the more
demanding SP~800-90B entropy-source assessment~\cite{nist80090b}, which
is the current NIST standard for qualifying entropy sources in
cryptographic applications.

This paper closes that gap. We characterize an MCP6002-based Zener TRNG
across five gain settings spanning nearly one decade ($G=11$ to $G=101$)
and a secondary coupling-capacitor study ($\Cg=0.1\,\upmu\mathrm{F}$
vs.\ $0.01\,\upmu\mathrm{F}$), subjecting each configuration to both
NIST~SP~800-22 and SP~800-90B in parallel. The experimental dataset uses
$1.44\!\times\!10^{9}$ raw bits per configuration at \SI{400}{\kilo\hertz} 
and is the largest published for this circuit family, enabling
statistical conclusions with confidence levels not achievable in
shorter captures.

The principal contributions of this work are:
\begin{itemize}
\item \textbf{Three-phase ACF structure.} A three-phase
positive--anti-correlated--independent autocorrelation structure
appears in every gain configuration, attributable to MCP6002
comparator saturation-recovery dynamics.

\item \textbf{Gain--entropy relationship.} SP~800-90B min-entropy
rises monotonically with gain across the $\Cg=0.1\,\upmu\mathrm{F}$
sweep ($\hmin=0.441$ at $G=11$ to $0.832$ at $G=101$), with the
compression estimator the binding constraint throughout.

\item \textbf{Entropy source attribution.} A DC substitution
experiment replacing the Zener diode with a stable
\SI{23.5}{\volt} DC source at the same Node~A bias point cuts mean
switching frequency by 97\% ($\SI{11841}{\hertz}\to\SI{407}{\hertz}$),
establishing Zener avalanche noise as the necessary entropy source
for this circuit architecture.

\item \textbf{Comparator-overdrive mechanism.} Node~C amplitude and
DC-floor offset govern the near-threshold oscillation rate, the
physical mechanism linking gain to entropy quality and bias
degradation.

\item \textbf{$\Cg$ dual role.} $\Cg=0.01\,\upmu\mathrm{F}$ achieves
committed NodeREF crossings and a Pareto improvement in entropy,
throughput, and bias over $\Cg=0.1\,\upmu\mathrm{F}$.

\item \textbf{Minimum viable gain boundary.} Convergent SP~800-22 and
SP~800-90B evidence identifies $G=48$ as optimal: 15/15 SP~800-22,
$\hmin=0.801$~b/b, IID pass, 11{,}568~b/s.
\end{itemize}

The remainder of this paper is organized as follows.
Section~\ref{sec:circuit} describes the circuit and apparatus.
Section~\ref{sec:method} details the data-processing methodology.
Section~\ref{sec:acf} presents the correlation-structure analysis.
Section~\ref{sec:overdrive} analyzes the comparator-overdrive mechanism.
Section~\ref{sec:cg} characterizes the $\Cg$ effect.
Sections~\ref{sec:nist22} and \ref{sec:nist90b} present the SP~800-22
and SP~800-90B results respectively. Section~\ref{sec:boundary}
establishes the minimum viable gain boundary.
Section~\ref{sec:ics} evaluates ICS applicability, and
Section~\ref{sec:conclusion} concludes.

\section{Circuit Design and Experimental Apparatus}\label{sec:circuit}
This section outlines the circuit design, experimental
configurations, and measurement instrumentation used throughout the
study.

\subsection{Circuit Architecture}

\begin{figure*}[t]
\centering
\includegraphics[width=0.96\textwidth, trim={2.1in 2.6in 2.1in 2.7in}, clip]{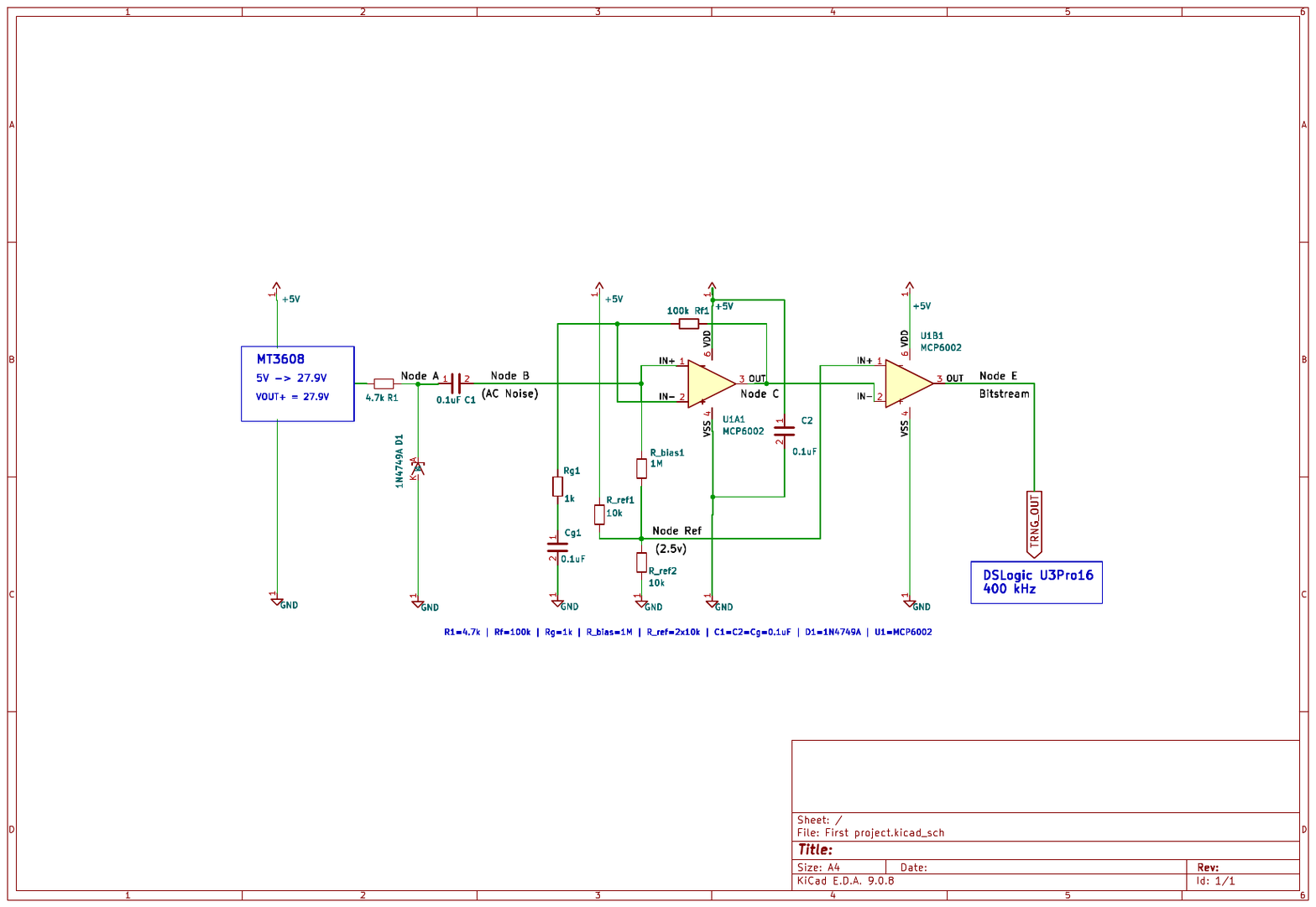}
\caption{Zener avalanche-noise TRNG schematic. An MT3608 boost
converter establishes \SI{27.9}{\volt} reverse bias on the 1N4749A
Zener through $R_b$, biasing Node~A at \SI{23.6}{\volt} in the
avalanche regime. The capacitor $C_g$ couples the noise into a
non-inverting amplifier (U1A) of gain $G=1+R_f/R_g$ that drives
Node~C. An open-loop comparator (U1B) thresholds Node~C against a
NodeREF of \SI{2.54}{\volt} produced by the $R_1$--$R_2$ divider.
The binary output at Node~E is captured by a DSLogic U3Pro16 logic
analyzer at \SI{400}{\kilo\hertz}. U1A and U1B are the two channels
of one MCP6002 dual op-amp.}
\label{fig:schematic}
\end{figure*}

The TRNG circuit comprises three functional stages: a regulated
high-voltage supply, a Zener avalanche-noise source with coupling
network, and a two-stage MCP6002 signal chain. The following description traces the signal path from the physical
noise source to the digital output at Node~E, shown schematically in
Fig.~\ref{fig:schematic}.

\textit{Stage~1 - High-voltage supply:} An MT3608 boost converter
steps up the \SI{5}{\volt} system supply to \SI{27.9}{\volt},
establishing the reverse-bias voltage for the Zener diode. The boost
converter output is stable to within $\pm 0.3$~V across all captures,
ensuring a consistent Zener operating point throughout the gain sweep.
Node~A-the voltage at the Zener anode, outputs \SI{23.6}{\volt}
under all operating conditions, confirming stable avalanche-regime
operation.

\textit{Stage~2 - Noise source and coupling:} We reverse bias a 1N4749A Zener diode
(\SI{24}{\volt} nominal) at Node~A. In the
avalanche-breakdown regime, the diode generates broadband shot noise
arising from the stochastic avalanche-multiplication process \cite{petrie2000}. The
noise current flows through the Zener impedance, developing a
small-signal voltage that constitutes the raw entropy input to the
amplifier stage. A coupling capacitor $\Cg$ in series with a gate
resistor $\Rg=\SI{1}{\kilo\ohm}$ forms an RC high-pass filter with
cutoff frequency $f_c = 1/(2\pi \Rg \Cg)$. In the primary study,
$\Cg=\SI{0.1}{\micro\farad}$ gives $f_c=\SI{1592}{\hertz}$;
a secondary study with $\Cg=\SI{0.01}{\micro\farad}$ raises $f_c$ to
\SI{15915}{\hertz}. We analyze the role of the filter in shaping both the noise
spectrum and the comparator dynamics in
Section~\ref{sec:cg}.

\textit{Stage~3 - Signal chain (MCP6002 dual op-amp):}
We configure the first op-amp section as a non-inverting voltage
amplifier with gain $G = 1 + \Rf/\Rg$. The feedback resistance $\Rf$ is
the primary experimental variable, taking values in
$\{10,22,30,47,100\}$~k$\Omega$ to produce gain settings
$G\!\in\!\{11,23,31,48,101\}$. The amplified noise appears at Node~C,
where its peak-to-peak voltage is characterized independently by an
Analog Discovery~3 (AD3) oscilloscope. We configure the second MCP6002 section as an open-loop voltage comparator, thresholding Node~C against a reference voltage NodeREF. The comparator output at Node~E
is the binary TRNG output captured by the DSLogic U3Pro16 logic
analyzer. We measure the NodeREF at \SI{2.54}{\volt} under operating
conditions, representing a $+\SI{40}{\milli\volt}$ offset above the
designed \SI{2.50}{\volt} midpoint. Section~\ref{sec:overdrive} analyzes the implications of the offset.

The MCP6002 has a 1~MHz gain--bandwidth product (GBW). At $G=101$,
the closed-loop bandwidth of the amplifier stage is
$\mathrm{GBW}/G \approx \SI{9.9}{\kilo\hertz}$-above the mean
switching frequency of \SI{11841}{\hertz}, ensuring that the amplifier
does not become the bandwidth bottleneck at the highest gain setting.
At $G=11$, the closed-loop bandwidth extends to approximately
\SI{91}{\kilo\hertz}, well beyond any relevant signal frequency. VDD
stays stable at 5.04--5.06~V across all captures. The complete circuit
fits on a standard breadboard with no specialized PCB or shielding.

\subsection{Experimental Configurations}

Table~\ref{tab:configs} summarizes the six capture configurations.
Runs~1--5 vary $\Rf$ while holding $\Cg=\SI{0.1}{\micro\farad}$
constant. Run~6 retains $\Rf=\SI{100}{\kilo\ohm}$ ($G=101$) and
reduces $\Cg$ to $\SI{0.01}{\micro\farad}$ for the coupling-capacitor
study. We compute the mean switching frequency $\favg$ from the
DSLogic capture as half the transition count divided by the capture
duration. Raw bias (ones fraction) is derived from the
$1.44\!\times\!10^{9}$ raw bits before any processing.
Sections~\ref{sec:acf}--\ref{sec:nist22} discuss observations from
these six configurations.

\begin{table}[t]
\caption{Experimental Configuration Parameters}
\label{tab:configs}
\centering\small
\begin{tabular}{@{}cccccrrc@{}}
\toprule
Run & $\Rf$ & $G$ & $\Cg$ & VDD & $\favg$ & Bias & Node~A \\
    & (k$\Omega$) &     & (\textmu{}F) & (V) & (Hz) & (\%) & (V) \\
\midrule
1 & 100 & 101 & 0.10 & 5.04 & 11{,}841 & 52.04 & 23.6 \\
2 &  47 &  48 & 0.10 & 5.05 & 14{,}449 & 52.50 & 23.6 \\
3 &  30 &  31 & 0.10 & 5.06 & 14{,}490 & 53.62 & 23.6 \\
4 &  22 &  23 & 0.10 & 5.05 & 14{,}137 & 55.43 & 23.6 \\
5 &  10 &  11 & 0.10 & 5.03 & 17{,}328 & 63.70 & 23.6 \\
6 & 100 & 101 & 0.01 & 5.05 & 15{,}580 & 51.48 & 23.6 \\
\bottomrule
\end{tabular}
\end{table}

\subsection{Measurement Instrumentation}

\textit{Digital output capture:} We use a DSLogic U3Pro16 logic analyzer
to sample Node~E at \SI{400}{\kilo\hertz} for 1~hour per configuration,
yielding $1.44\!\times\!10^{9}$ raw bits. The U3Pro16 captures all
comparator transitions including rapid chatter events that analog
instruments may miss.

\textit{Analog characterization:} The Analog Discovery~3 (AD3) measures
Node~C peak-to-peak voltage, maximum, minimum, and mean level under
continuous noise-driven operation. The AD3 also measures Node~E
frequency via trigger counting; however, a systematic discrepancy of
15--61\% was observed between AD3 and DSLogic frequency estimates
because the analog trigger misses rapid chatter transitions. All
frequency values cited throughout this paper use DSLogic measurements
as the authoritative reference.

\section{Data Processing Methodology}\label{sec:method}
This section outlines the data processing methodology and discusses the configurations for the NIST SP 800-22, and 800-90B assessment protocol.

\subsection{Decimation and Independence Criterion}

Raw \SI{400}{\kilo\hertz} captures exhibit inter-sample
correlation: at a mean switching frequency of 11--17~kHz, the binary
output changes state only once per 23--36 samples on average. Adjacent
raw samples are therefore identical with probability exceeding 0.97.
Applying statistical tests directly to the raw bitstream would
characterize the sampling clock rather than the entropy source.

Decimation by factor $D=120$ maps the \SI{400}{\kilo\hertz} capture to
an effective rate of \SI{3333}{\hertz}, corresponding to a physical
inter-sample spacing of $\Delta t = D/f_s = \SI{300}{\micro\second}$.
The choice of $D=120$ is justified empirically by the autocorrelation
analysis in Section~\ref{sec:acf}: across all five
$\Cg=\SI{0.1}{\micro\farad}$ configurations, the Autocorrelation Function (ACF) magnitude falls
below $|r|<0.005$ by \SI{290}{\micro\second}. Retained samples at
\SI{300}{\micro\second} spacing thus reside in the statistically
independent regime for every configuration. After decimation,
$N_{\text{dec}} = 12\!\times\!10^{6}$ bits are available per
configuration, packing to $1.50\!\times\!10^{6}$ bytes and satisfying
the SP~800-90B minimum of $10^{6}$ bytes.

\subsection{Double Von Neumann Debiasing}

Raw decimated bitstreams exhibit positive bias ranging from 52.0\% to
63.7\% ones (Table~\ref{tab:configs}), arising from the NodeREF offset
and DC-floor dynamics detailed in Section~\ref{sec:overdrive}. The
von~Neumann (VN) extractor~\cite{vonneumann1951} processes
non-overlapping consecutive bit pairs: discordant pairs
($01\!\to\!0$, $10\!\to\!1$) are output; concordant pairs ($00$, $11$)
are discarded. This procedure removes first-order bias with output
efficiency $\eta = 2p(1-p)$, where $p$ is the ones fraction.

A single VN pass reduces bias to 50.01--50.03\% but leaves residual
serial correlation $|r|\!\approx\!0.003$--$0.005$ in the output
bitstream. At $n=12\!\times\!10^{6}$ bits, this correlation level is
sufficient to cause statistically significant failures in SP~800-22
T03 (Runs), T11 (Serial), and T12 (Approximate Entropy). A second VN
pass (Double~VN, DVN) reduces correlation to $|r|<0.003$ at the cost
of approximately 75\% of the single-pass output, yielding
$N_{\text{dvn}}\!\in\![694\mathrm{k}, 751\mathrm{k}]$ bits per
configuration, sufficient for the full fifteen-test SP~800-22 suite.
We apply DVN uniformly across all six configurations to ensure
methodological consistency.

\subsection{NIST SP 800-22 Assessment Protocol}

We implement fifteen SP~800-22 tests in Python,
and validate each against the reference $p$-values in SP~800-22
Appendix~A. The T10 (Linear Complexity) test employed a Berlekamp--Massey
algorithm over GF(2) capped at $N=500$ blocks of $M=500$ bits to avoid
memory exhaustion, validated on analytically verified test vectors
(all-zeros, all-ones, and alternating sequences with analytically
known LFSR lengths). All fifteen tests are run on the DVN output for
every configuration. The Python code evaluates the T14 (Random Excursions) and T15 (Excursions
Variant) on applicability criteria
($J\geq 500$ zero crossings) prior to execution.

\subsection{NIST SP 800-90B Assessment Protocol}

We apply the SP~800-90B bitstring estimator path to the full
$N_{\text{dec}}=12\!\times\!10^{6}$ decimated bits (prior to VN
debiasing), packed into $1.50\!\times\!10^{6}$ bytes. Using the validated 
\texttt{sp800-90b} Python package, the pipeline executes the ten bitstring
estimators:
Most~Common~Value (MCV), Collision, Markov, Compression, T-Tuple,
Longest~Repeated~Substring (LRS), Multi-MCV, Multi-Markov, and the
binding minimum~$\hmin$. IID assessment used the SP~800-90B
permutation-test procedure (Section~4.3 of the standard). All
estimates are reported in bits per bit (b/b), obtained by dividing
the package per-byte output by~8.

\section{Correlation Structure Analysis}\label{sec:acf}

\begin{figure}[t]
\centering
\includegraphics[width=\columnwidth]{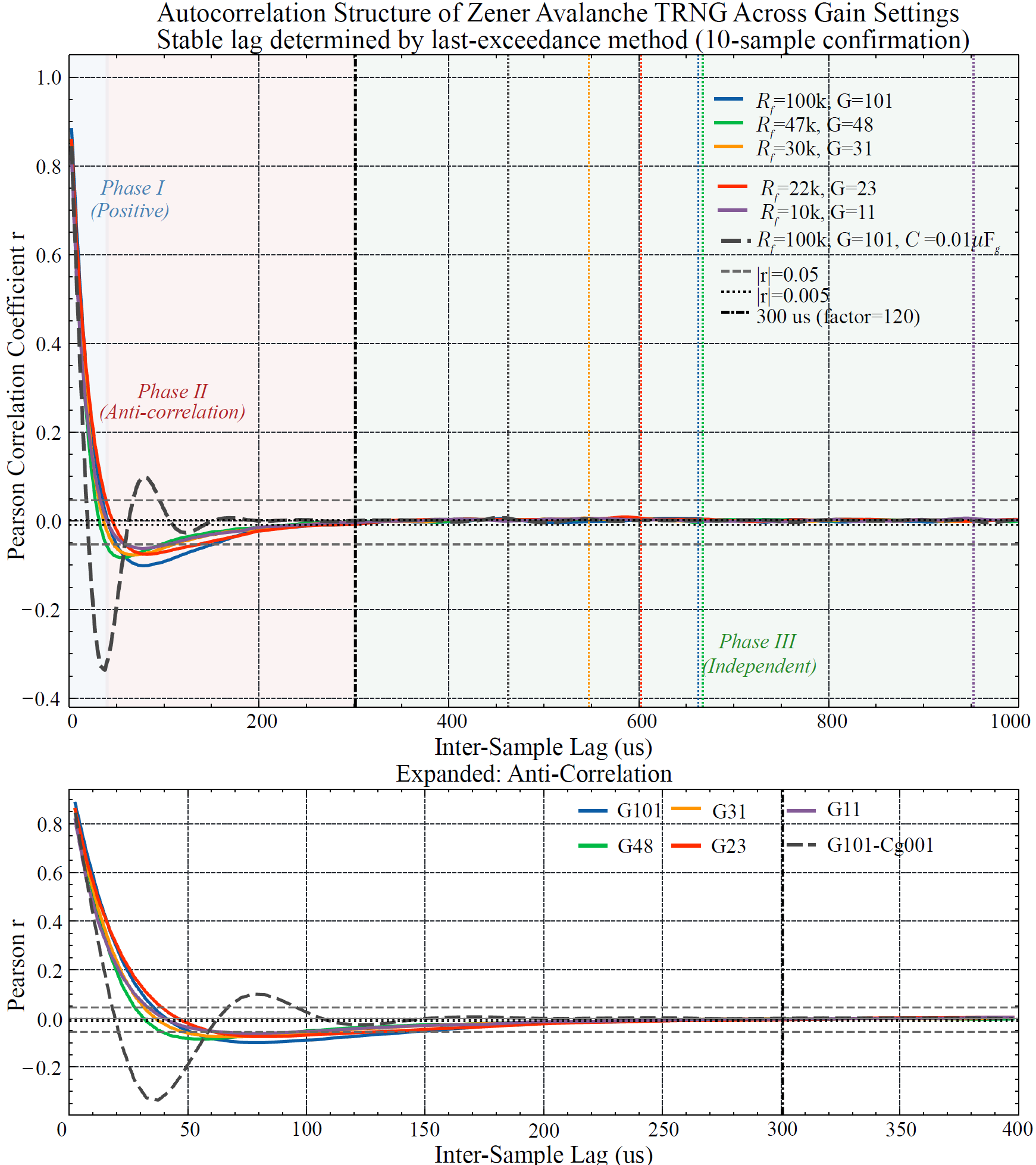}
\caption{Sample autocorrelation function of the raw \SI{400}{\kilo\hertz}
Node~E bitstream as a function of physical time lag (\textmu{}s).
Top: full lag range showing the three-phase structure-Phase~I
(positive), Phase~II (anti-correlation), Phase~III (independent)-%
common to all five $C_g\!=\!\SI{0.1}{\micro\farad}$ gain
configurations. The vertical dash-dotted line marks the
\SI{300}{\micro\second} decimation spacing ($D=120$). The dashed
trace ($C_g\!=\!\SI{0.01}{\micro\farad}$) exhibits substantially
deeper anti-correlation. Bottom: expanded view of the 0--400~\textmu{}s
anti-correlation zone.}
\label{fig:acf}
\end{figure}

This section outlines the autocorrelation structure of Zener Avalanche TRNG across gain setting and determines the best sampling lag.
\subsection{Three-Phase ACF Structure}

Computation of the sample autocorrelation function on the raw
\SI{400}{\kilo\hertz} bitstream reveals a three-phase structure that
is consistent across all five $\Cg=\SI{0.1}{\micro\farad}$
gain configurations. 

\textit{Phase~I - Positive correlation (0--40~\textmu{}s):} Within
each switching half-cycle, the binary output maintains its current
state, producing positive ACF for lags shorter than the mean
half-period. All five configurations exhibit positive ACF out to
approximately \SI{40}{\micro\second}, consistent with the 11--17~kHz
mean switching frequencies observed across gain settings.

\textit{Phase~II - Anti-correlation (40--290~\textmu{}s):} A
negative ACF lobe follows the positive phase. We attribute
this anti-correlation to MCP6002 comparator saturation recovery. When
the amplified noise signal drives the comparator output to saturation,
the recovery trajectory produces a transient overshoot in the opposite
direction. This overshoot is reflected in the probability that the
next sample takes the opposite value, manifesting as negative ACF.
The depth of this phase scales with gain (Table~\ref{tab:acf}),
consistent with deeper saturation at higher amplification levels.

\textit{Phase~III - Independence ($>290$~\textmu{}s):} The ACF
magnitude decays below $|r|=0.005$ for all configurations by
\SI{290}{\micro\second}, entering the statistically independent
regime (Fig.~\ref{fig:acf}, top panel). Accordingly, we select a
decimation spacing of \SI{300}{\micro\second}, which provides
independent values for all gain settings.

\begin{table}[t]
\caption{Autocorrelation Structure Parameters Across All Gain Configurations}
\label{tab:acf}
\centering\small
\setlength{\tabcolsep}{3.5pt}
\begin{tabular}{@{}rrrrrrrr@{}}
\toprule
Config              & $\favg$  & Bias  & Peak~$r$ & Peak   & Stable & Decorr. \\
($G\!=\!$)              & (Hz)     & (\%)  &          & Lag (\textmu{}s) & Lag (\textmu{}s)& Lag (\textmu{}s) \\
\midrule
$101$         & 11{,}841 & 52.08 & $-0.101$ & 77.5  & 662.5 & 37.5 \\
$48$          & 14{,}449 & 52.46 & $-0.085$ & 57.5  & 667.5 & 27.5 \\
$31$          & 14{,}490 & 53.74 & $-0.077$ & 67.5  & 547.5 & 32.5 \\
$23$          & 14{,}137 & 55.58 & $-0.076$ & 82.5  & 602.5 & 42.5 \\
$11$          & 17{,}328 & 63.68 & $-0.063$ & 77.5  & 952.5 & 32.5 \\
$101$,  & 15{,}580 & 51.46 & $-0.337$ & 37.5 & 462.5 & 62.5 \\
$0.01$\textmu{}F \\
\bottomrule
\end{tabular}
\end{table}

\subsection{Gain Dependence of Anti-Correlation Depth}

Anti-correlation depth (peak negative $r$) decreases monotonically as
gain decreases: from $r=-0.101$ at $G=101$ to $r=-0.063$ at $G=11$.
This ordering reflects the saturation-recovery mechanism. Higher gain
drives the MCP6002 output deeper into rail saturation, producing a
longer and more energetic recovery overshoot. Notably, deeper
anti-correlation does \emph{not} imply lower entropy---the compression
estimator results in Section~\ref{sec:nist90b} show that
configurations with shallower anti-correlation (lower gain) have \emph{lower} $\hmin$, not higher, because bit bursts arising from near-threshold oscillation are the dominant entropy-limiting mechanism at low gain.

\subsection{$\Cg$ Effect on ACF}

Replacing $\Cg=\SI{0.1}{\micro\farad}$ with
$\Cg=\SI{0.01}{\micro\farad}$ at $G=101$ increases the
anti-correlation depth by $3.3\times$ (peak $r=-0.337$ vs.\ $-0.101$)
and reduces the stable independence lag from \SI{663}{\micro\second}
to \SI{463}{\micro\second}. The deeper anti-correlation reflects the
larger Node~C amplitude (\SI{158.6}{\milli\volt} vs.\
\SI{51.3}{\milli\volt}) driving stronger comparator saturation and
recovery ringing. The shorter independence lag is consistent with the
higher switching frequency (\SI{15580}{\hertz} vs.\
\SI{11841}{\hertz}) enabled by the reduced capacitive loading on the
amplifier input. Both configurations satisfy the
\SI{300}{\micro\second} independence criterion, and the decimation
factor $D=120$ remains valid for all six configurations.

\section{Comparator Overdrive Mechanism}\label{sec:overdrive}

\begin{figure*}[t]
\centering
\includegraphics[width=0.95\textwidth]{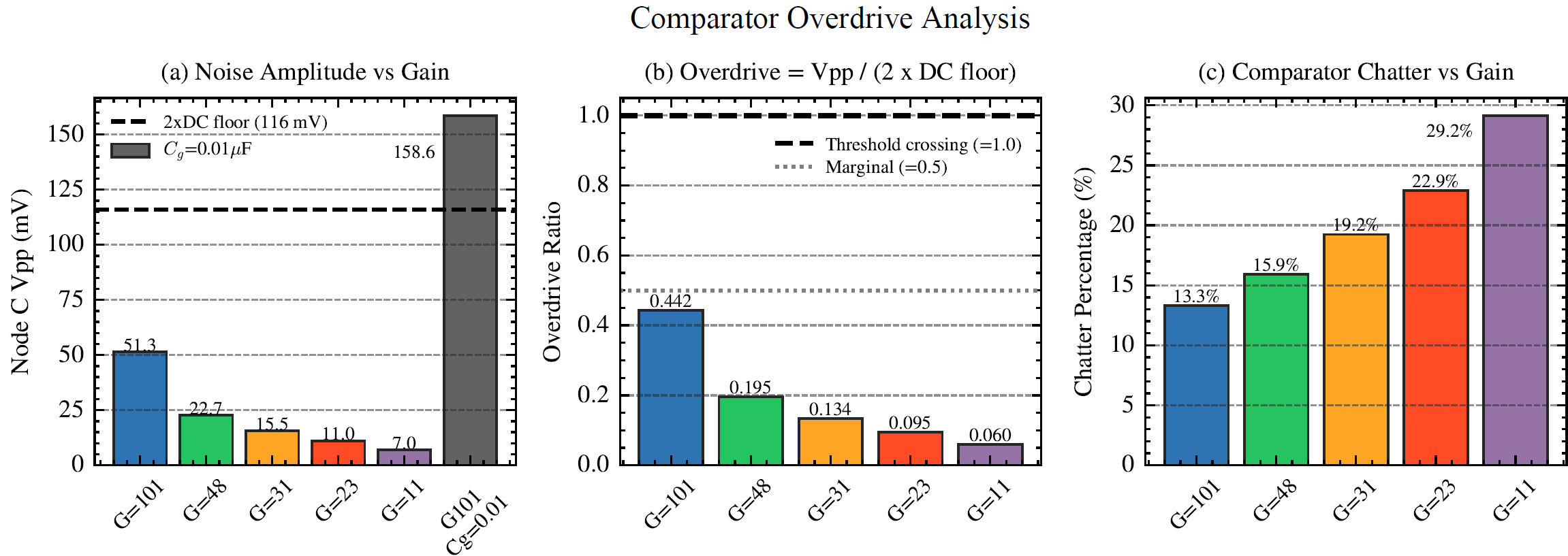}
\caption{Comparator overdrive analysis across gain configurations.
\textbf{(a)} Node~C peak-to-peak amplitude (mV) measured by AD3.
The dashed horizontal line marks $2\!\times\!\mathrm{DC\,floor}=\SI{116}{\milli\volt}$,
the threshold for committed NodeREF crossing. Only $C_g\!=\!\SI{0.01}{\micro\farad}$
exceeds this threshold; all five $C_g\!=\!\SI{0.1}{\micro\farad}$ configurations
fall well below it. \textbf{(b)} Overdrive ratio
$V_{pp}/(2\!\times\!\mathrm{DC\,floor})$ for the $C_g\!=\!\SI{0.1}{\micro\farad}$
sweep, decreasing monotonically from 0.44 ($G\!=\!101$) to 0.06 ($G\!=\!11$).
\textbf{(c)} Near-threshold oscillation (NTO) rate, defined as the
fraction of transitions arising from sub-committed comparator
perturbation rather than genuine noise-driven crossings. The NTO rate
increases monotonically as gain decreases, from 13.3\% at $G\!=\!101$
to 29.2\% at $G\!=\!11$.}
\label{fig:overdrive}
\end{figure*}
This section analyzes the comparator threshold geometry and overdrive mechanism, confirming that the observed noise originates from Zener avalanche dynamics rather than MCP6002 self-excitation.
\subsection{DC Floor and Threshold Geometry}

The measured NodeREF of \SI{2.54}{\volt} ($+\SI{40}{\milli\volt}$
above design) and a measured DC floor of \SI{58}{\milli\volt}
establish the threshold geometry for the comparator. Node~C must swing
to at least $(\text{NodeREF}-\SI{58}{\milli\volt})=\SI{2.482}{\volt}$
to produce a committed low-to-high transition at Node~E.
Table~\ref{tab:overdrive} shows that for all five
$\Cg=\SI{0.1}{\micro\farad}$ configurations, Node~C minimum remains at
47.6--60.8~mV \emph{above} NodeREF. The signal never crosses the
threshold. We therefore characterize all observed binary switching at
$\Cg=\SI{0.1}{\micro\farad}$ as \emph{near-threshold oscillation
(NTO)}---Zener noise perturbing Node~C close enough to NodeREF to
repeatedly destabilize the comparator without producing a committed
crossing. Section~\ref{sec:dcsub} verifies experimentally that this
oscillation requires the Zener noise and does not arise from MCP6002
self-excitation.

With $\Cg=\SI{0.01}{\micro\farad}$, Node~C minimum reaches
$-\SI{14.8}{\milli\volt}$ relative to NodeREF, achieving committed
threshold crossing by a margin of \SI{72.8}{\milli\volt}. This single
component change qualitatively alters the entropy-generation
mechanism from \textit{near-threshold Zener-perturbed oscillation to
committed crossing-driven switching} with consequences for both
entropy quality and waveform fidelity discussed in
Section~\ref{sec:cg}.

\begin{table}[t]
\caption{Node~C Amplitude, Overdrive, and Near-Threshold Oscillation Analysis (AD3 Measurements)}
\label{tab:overdrive}
\centering\small
\setlength{\tabcolsep}{3.7pt} 

\begin{threeparttable} 
\begin{tabular}{@{}lrrrrrr@{}}
\toprule
Config & Vpp & Max & Min & OD & NTO & $f_{\text{avg}}$ \\
       & (mV) & (mV)& (mV)&    & (\%)    & (Hz) \\
\midrule
$G\!=\!101$, $0.1$\textmu{}F  & 51.3   & $+98.9$ & $+47.6$ & 0.44 & 13.3 & 11{,}841 \\
$G\!=\!48$, $0.1$\textmu{}F   & 22.7   & $+80.6$ & $+58.0$ & 0.20 & 15.9 & 14{,}449 \\
$G\!=\!31$, $0.1$\textmu{}F   & 15.5   & $+73.3$ & $+57.7$ & 0.13 & 19.2 & 14{,}490 \\
$G\!=\!23$, $0.1$\textmu{}F   & 11.0   & $+69.2$ & $+58.2$ & 0.09 & 22.9 & 14{,}137 \\
$G\!=\!11$, $0.1$\textmu{}F   & 6.96   & $+67.7$ & $+60.8$ & 0.06 & 29.2 & 17{,}328 \\
$G\!=\!101$, $0.01$\textmu{}F & 158.6  & $+143.8$& $-14.8$ & 1.37 & 7.9  & 15{,}580 \\
\bottomrule
\end{tabular}

\begin{tablenotes}
  \footnotesize
  \item[] OD = overdrive = $V_{pp}/(2\!\times\!\mathrm{DC\,floor})$;\\ NTO = near-threshold oscillation rate.
\end{tablenotes}
\end{threeparttable}

\end{table}

\subsection{Near-Threshold Oscillation Rate and Bias Degradation}

Although Node~C does not achieve committed threshold crossings in the
$\Cg=\SI{0.1}{\micro\farad}$ configurations, Zener avalanche noise
perturbs Node~C sufficiently near NodeREF to destabilize the
comparator repeatedly. The near-threshold oscillation (NTO)
rate increases
monotonically from 13.3\% at $G=101$ to 29.2\% at $G=11$
(Table~\ref{tab:overdrive}, Fig.~\ref{fig:overdrive}c), reflecting the
decreasing Node~C amplitude with decreasing gain
(Fig.~\ref{fig:overdrive}a). At the lowest gain setting $G=11$, the \SI{6.96}{\milli\volt}
swing is barely above the measurement noise floor.

The raw bitstream bias degrades in direct proportion to the NTO rate.
At $G=11$, the 63.7\% ones fraction arises from an asymmetry in the
MCP6002 settling behavior during the indeterminate near-threshold
period. The output spends more time at the positive rail than the
negative rail. This bias is deterministic and removable,
Double~VN debiasing restores all configurations to 49.95--50.11\%
within the VN output, demonstrating that the bias reflects a
systematic circuit asymmetry rather than a fundamental entropy
deficit.

\subsection{Gain to Entropy Quality Relationship}

The physical chain linking gain to entropy quality operates as
follows. Higher $\Rf$ increases Node~C amplitude. Larger amplitude
provides more decisive comparator perturbation, reducing the NTO
fraction and producing a more symmetric bit distribution.
Simultaneously, larger amplitude drives the MCP6002 deeper into
saturation on each switching event, lengthening the recovery
trajectory and producing deeper Phase~II anti-correlation in the ACF.
Lower gain inverts both effects. The small Node~C swing perturbs the
comparator weakly, generating dense near-threshold bit bursts that
create correlated sequences and, at $G=11$, ultimately cause IID
failure under SP~800-90B permutation testing. The $\Cg$ parameter
modulates this mechanism by controlling both the noise bandwidth
reaching the amplifier input and, critically, whether the resulting Node~C amplitude is sufficient to achieve committed crossings, as analyzed in Section~\ref{sec:cg}.

\subsection{Entropy Source Verification: DC Substitution Test}
\label{sec:dcsub}

To verify experimentally whether the near-threshold oscillation
described above requires Zener avalanche noise, or could instead arise
from MCP6002 self-excitation alone, we conducted a DC substitution
experiment. We removed the 1N4749A Zener diode and retuned the
existing MT3608 boost converter to \SI{23.5}{\volt}, connecting its
output directly to Node~A through a \SI{100}{\micro\farad}
electrolytic capacitor and a \SI{100}{\nano\farad} ceramic capacitor
(suppressing the converter's \SI{1.2}{\mega\hertz} switching ripple)
and a $\SI{10}{\kilo\ohm}$ dummy load (maintaining stable
boost-converter regulation in the absence of the Zener's current
draw). We verified Node~A ripple below \SI{5}{\milli\volt}
peak-to-peak with the AD3 at \SI{1}{\micro\second}/div before each
capture. All other circuit parameters such as gain, $\Cg$, NodeREF, and
DSLogic capture settings remained identical to the original
captures.

With $\Cg=\SI{0.1}{\micro\farad}$ and $G=101$, DC substitution reduced
 the mean switching frequency from \SI{11841}{\hertz}
 (Table~\ref{tab:configs}) to approximately \SI{407}{\hertz}, a
 reduction of approximately 97\%. Node~E remained predominantly high,
 with rare, irregularly spaced dips consistent with environmental noise
 threshold crossings rather than self-sustaining comparator
 oscillation.

This result yields three conclusions. First, MCP6002 internal noise
alone is insufficient to sustain comparator switching at this
topology's operating point. The DC-floor geometry places NodeREF too
close to Node~C's resting potential for thermal noise to bridge the
gap reliably. Second, Zener avalanche noise is the necessary entropy
source for all six configurations, including those at
$\Cg=\SI{0.1}{\micro\farad}$ where Node~C does not achieve a committed
crossing. Third, the distinction between the two $\Cg$ regimes is one
of \emph{overdrive quality} rather than \emph{entropy-source
identity}. In both regimes the Zener noise is essential, but
$\Cg=\SI{0.01}{\micro\farad}$ converts this noise into committed
threshold crossings, whereas $\Cg=\SI{0.1}{\micro\farad}$ produces
near-threshold oscillation that is statistically weaker yet
fundamentally Zener-driven.

\section{Coupling Capacitor Effect Study}\label{sec:cg}
This section characterizes the dual filtering and stabilization roles
of the coupling capacitor $\Cg$ and quantifies its impact on entropy
quality and throughput at $G=101$.

\begin{figure}[t]
\centering
\includegraphics[width=\columnwidth]{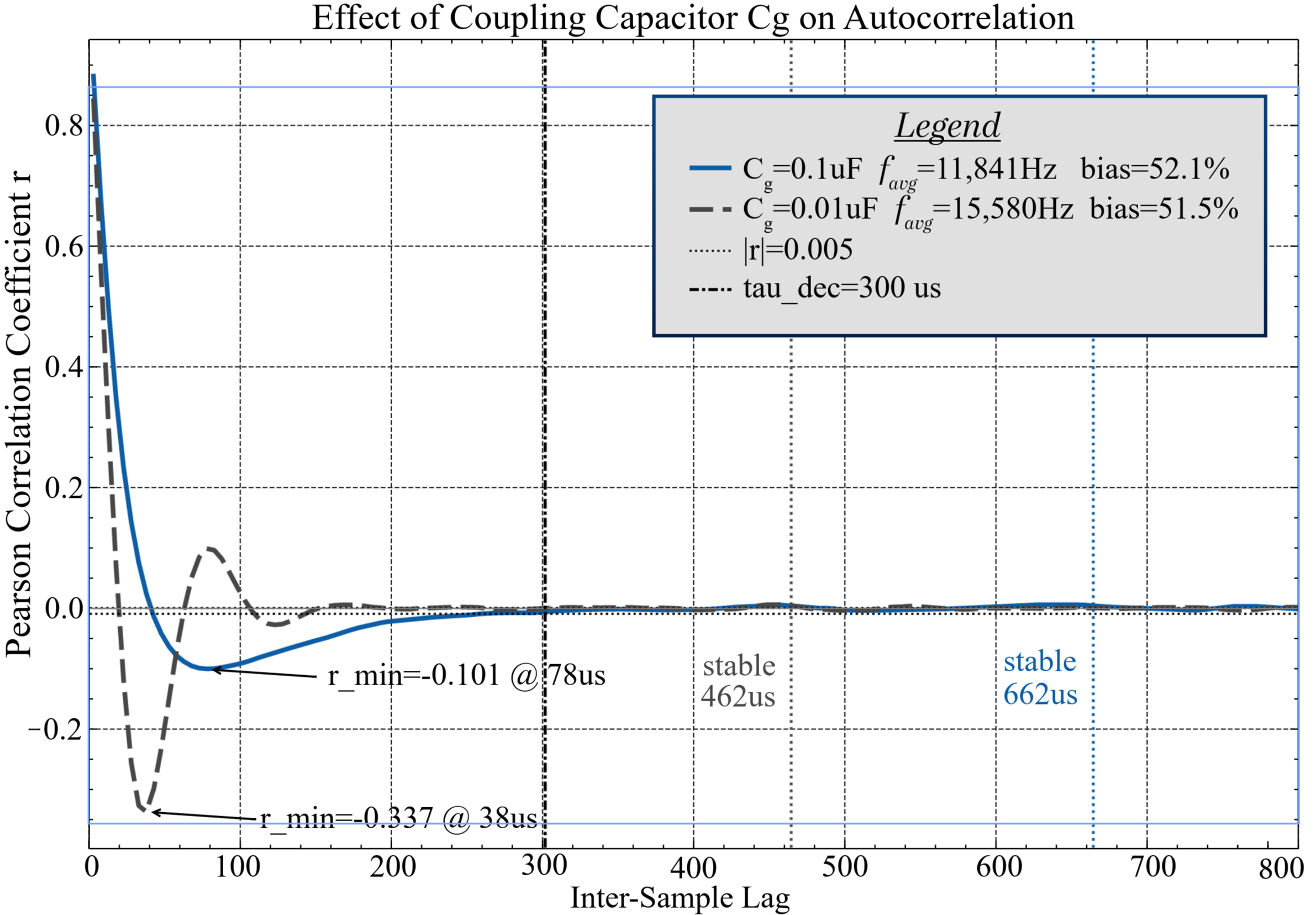}
\caption{Effect of coupling capacitor on the autocorrelation
function at $G=101$ ($R_f\!=\!\SI{100}{\kilo\ohm}$). Reducing $C_g$ from
\SI{0.1}{\micro\farad} (solid) to \SI{0.01}{\micro\farad} (dashed)
deepens the Phase~II anti-correlation by $3.3\times$
($r_{\min}\!=\!-0.337$ vs.\ $-0.101$), shortens the peak ringing
lag (\SI{38}{\micro\second} vs.\ \SI{78}{\micro\second}), and
shortens the stable independence lag from \SI{662}{\micro\second}
to \SI{462}{\micro\second}. Both configurations achieve $|r|<0.005$
by the \SI{300}{\micro\second} decimation threshold (vertical
dash-dotted line).}
\label{fig:cg_acf}
\end{figure}

\subsection{Dual Role of $\Cg$}

The coupling capacitor $\Cg$ in series with $\Rg$ forms an RC
high-pass filter with cutoff frequency $f_c=1/(2\pi \Rg \Cg)$. With
$\Rg=\SI{1}{\kilo\ohm}$ and $\Cg=\SI{0.1}{\micro\farad}$, $f_c=\SI{1592}{\hertz}$ stays below the mean switching frequency $\favg=\SI{11841}{\hertz}$. The capacitor therefore attenuates the
noise signal before it reaches the amplifier input, suppressing
spectral content above \SI{1592}{\hertz} and limiting the Node~C
amplitude to \SI{51.3}{\milli\volt} despite the high gain of 101.
With $\Cg=\SI{0.01}{\micro\farad}$, $f_c=\SI{15915}{\hertz}$
slightly exceeds $\favg=\SI{15580}{\hertz}$, allowing the full noise
bandwidth to reach the amplifier. This filter role together with the Zener-driven entropy source verified in Section~\ref{sec:dcsub} explains the
$3.1\times$ increase in $V_{pp}$ and the 31.6\% increase in switching
frequency when $\Cg$ is reduced.

$\Cg$ also plays a second, less obvious role as a comparator-%
stabilization element. By limiting the rate of change of the signal
at the amplifier input, the larger $\Cg=\SI{0.1}{\micro\farad}$ slows
the transitions at Node~C, reducing the comparator's tendency toward
  near-threshold oscillation at the transition edges. With
$\Cg=\SI{0.01}{\micro\farad}$, the faster signal transitions and
larger amplitude produce genuine threshold crossings at the cost of
deeper ACF anti-correlation (peak $r=-0.337$ vs.\ $-0.101$,
Fig.~\ref{fig:cg_acf}), reflecting the more energetic saturation
and recovery dynamics associated with clean crossing events.

\subsection{Impact on Entropy Quality}

Table~\ref{tab:cg} summarizes the $\Cg$ comparison at $G=101$.
Despite its deeper anti-correlation,
$\Cg=\SI{0.01}{\micro\farad}$ achieves higher $\hmin$
(0.869~b/b vs.\ 0.832~b/b) under SP~800-90B. The compression estimator's sensitivity rewards the faster alternation rate and reduced near-threshold oscillation of genuine crossing events, while the anti-correlation occurs on a timescale (37--463~\textmu{}s) below the \SI{300}{\micro\second} decimation threshold. Entropy throughput $\Phi=\favg\!\times\!\hmin$ reaches
13{,}534~b/s with $\Cg=\SI{0.01}{\micro\farad}$, a 37\% improvement
over the 9{,}856~b/s achieved with $\Cg=\SI{0.1}{\micro\farad}$ at
the same gain setting.

\begin{table}[t]
\caption{$\Cg$ Effect Summary at $G=101$ ($\Rf=\SI{100}{\kilo\ohm}$)}
\label{tab:cg}
\centering\small
\begin{tabular}{@{}lrrr@{}}
\toprule
Metric & $\Cg\!=\!0.1\,\upmu$F & $\Cg\!=\!0.01\,\upmu$F & Change \\
\midrule
$f_c$ (Hz)              & 1{,}592   & 15{,}915  & $\times 10.0$  \\
$V_{pp}$ at Node~C (mV) & 51.3      & 158.6     & $\times 3.09$  \\
Crosses NodeREF         & No        & Yes       & ---            \\
NTO (\%)            & 13.3      & 7.9       & $-40\%$        \\
$\favg$ (Hz)            & 11{,}841  & 15{,}580  & $+31.6\%$      \\
Raw bias (\%)           & 52.04     & 51.48     & $-0.56\%$      \\
ACF peak $r$            & $-0.101$  & $-0.337$  & $3.3\times$ deeper \\
$\hmin$ (b/b)           & 0.832     & 0.869     & $+0.037$       \\
Throughput (b/s)        & 9{,}856   & 13{,}534  & $+37\%$        \\
\bottomrule
\end{tabular}
\end{table}

These results establish that for this circuit topology, reducing
$\Cg$ from \SI{0.1}{\micro\farad} to \SI{0.01}{\micro\farad}
represents a Pareto improvement; entropy quality, throughput, and
bias all improve simultaneously. The trade-off is a deeper ACF
anti-correlation that does not survive decimation and therefore does
not affect the assessed entropy. For implementations where the single
passive-component substitution is feasible,
$\Cg=\SI{0.01}{\micro\farad}$ is unambiguously preferred.

\section{NIST SP 800-22 Statistical Test Suite Results}\label{sec:nist22}
This section presents the NIST SP~800-22 fifteen-test results across
all six configurations and analyzes the marginal $p$-values observed
at lower gain settings.
\subsection{Summary of Results}

\begin{table}[t]
\caption{NIST SP 800-22 Results (Double VN Debiasing, $\alpha=0.01$, $n\!\in\![694\mathrm{k}, 751\mathrm{k}]$ bits)}
\label{tab:nist22}
\centering\small
\setlength{\tabcolsep}{3.4pt} 
\begin{tabular}{@{}lcccccccr@{}}
\toprule
Gain Config & T03 & T06 & T11 & T13 & T14 & T15 & Pass & Overall \\
\midrule
$101$, $0.1$\textmu{}F & \checkmark & \checkmark & \checkmark & \checkmark & 0.008$^{\dagger}$ & \checkmark & 14/15 & MARG. \\
$48$, $0.1$\textmu{}F  & \checkmark & \checkmark & \checkmark & \checkmark & 0.048             & \checkmark & 15/15 & PASS$^{\star}$ \\
$31$, $0.1$\textmu{}F  & 0.010$^{\dagger}$ & \checkmark & \checkmark & \checkmark & ---$^{\ddagger}$ & ---$^{\ddagger}$ & 12/13 & MARG. \\
$23$, $0.1$\textmu{}F  & 0.008$^{\dagger}$ & \checkmark & \checkmark & \checkmark & 0.005$^{\dagger}$ & \checkmark & 13/15 & MARG. \\
$11$, $0.1$\textmu{}F  & \checkmark & \checkmark & \checkmark & \checkmark & ---$^{\ddagger}$ & ---$^{\ddagger}$ & 13/13 & PASS \\
$101$, $0.01$\textmu{}F & \checkmark & \checkmark & \checkmark & \checkmark & ---$^{\ddagger}$ & ---$^{\ddagger}$ & 13/13 & PASS \\
\bottomrule
\multicolumn{9}{l}{\footnotesize \checkmark\ PASS ($p\geq 0.01$). $^{\dagger}$Marginally below $\alpha$; $^{\star}$Sole 15/15 configuration.}\\
\multicolumn{9}{l}{\footnotesize $^{\ddagger}$T14/T15 not applicable ($J<500$ zero crossings).}\\
\end{tabular}
\end{table}

Table~\ref{tab:nist22} presents the SP~800-22 results on the Double~VN
output for each configuration. We execute fifteen tests for each of
the six configurations. With fifteen simultaneous tests per
configuration at $\alpha=0.01$, a genuinely random source has
probability $P=1-(0.99)^{15}\approx13.99\%$ of producing at least one
marginal result by chance alone, so the four marginal $p$-values
observed across the dataset are consistent with expected
multiple-testing variation rather than a systematic entropy defect.
The T14 results at $G=101$ ($p=0.008$) and $G=23$
($p=0.005$) are due to multiple-testing statistical
fluctuation.

The T13 (Cumulative Sums) test, which is sensitive to
the same walk drift that causes T14 failures, passes at all
configurations ($p$ ranging from 0.21 to 0.84), ruling out systematic
walk asymmetry as the cause. The T03 marginal failures at $G=31$
($p=0.010$) and $G=23$ ($p=0.008$) are attributable to residual
serial correlation $|r|\!\approx\!0.003$ in the Double~VN output at
these higher-bias configurations. At $n\!\approx\!745\mathrm{k}$
bits, T03 has statistical power to detect correlations of this
magnitude. Both affected configurations achieve $\hmin>0.67$~b/b
under SP~800-90B, confirming that the T03 sensitivity at this sample
size reflects a processing artifact rather than a fundamental entropy
deficit.

T14 and T15 inapplicability ($J<500$ zero crossings) at $G=31$,
$G=11$, and the $\Cg=\SI{0.01}{\micro\farad}$ configuration arises
from the combination of DVN output size ($\sim 750\mathrm{k}$ bits)
and residual bias below 0.05\%. The random walk from these
configurations does not return to zero frequently enough to satisfy
the T14 applicability criterion. All three affected configurations
pass all applicable tests.

\subsection{$G=48$ as the Optimal Configuration}

$G=48$ ($\Rf=\SI{47}{\kilo\ohm}$, $\Cg=\SI{0.1}{\micro\farad}$) is
the sole configuration achieving a perfect 15/15 score. T14
($p=0.048$) and T15 ($p=0.031$) both pass with substantial margin,
confirming symmetric random-walk behavior after debiasing. This
result, combined with the SP~800-90B findings in
Section~\ref{sec:nist90b}, provides a convergent identification of
$G=48$ as the optimal operating point by two independent assessment
methodologies.

\section{NIST SP 800-90B Entropy Source Assessment}\label{sec:nist90b}
This section presents the NIST SP~800-90B entropy source assessment,
identifies the binding min-entropy estimator, and analyzes IID
compliance and entropy throughput across configurations.

\begin{figure*}[t]
\centering
\includegraphics[width=0.95\textwidth]{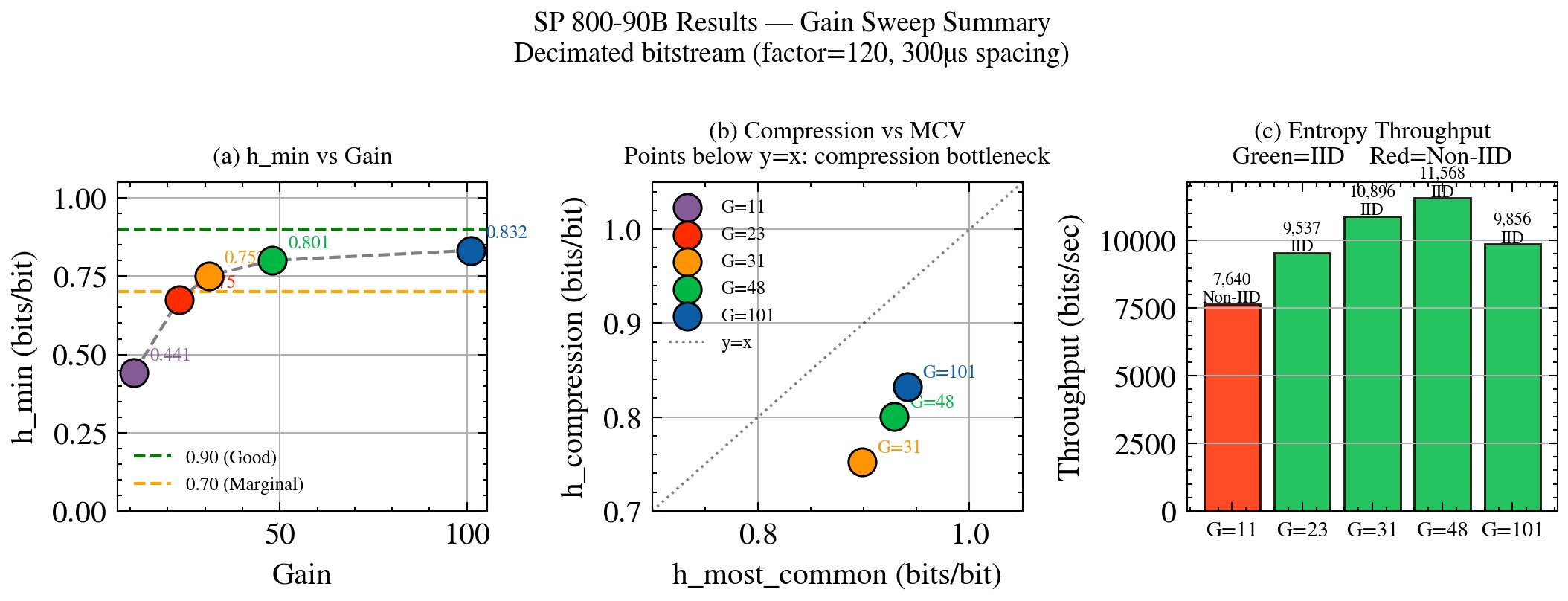}
\caption{NIST SP 800-90B assessment summary on the decimated
bitstream ($D=120$, \SI{300}{\micro\second} spacing).
\textbf{(a)} Min-entropy $h_{\min}$ versus gain. Green markers
indicate IID pass; red marker (G=11) indicates IID failure under
the SP~800-90B permutation test. The \emph{Marginal} threshold
($h_{\min}=0.7$) is marked. \textbf{(b)} Compression-estimator
output $h_{\mathrm{compression}}$ plotted against
most-common-value $h_{\mathrm{MCV}}$. All points fall well below
the $y\!=\!x$ line, demonstrating that the compression estimator is
the universally binding constraint
($h_{\min}=h_{\mathrm{compression}}$ for every configuration).
\textbf{(c)} Entropy throughput $\favg\!\times\!\hmin$ in bits/s,
peaking at 11{,}568~b/s for $G=48$ within the
$\Cg=\SI{0.1}{\micro\farad}$ sweep.}
\label{fig:sp80090b}
\end{figure*}

\subsection{Min-Entropy Results and Compression Bottleneck}

Table~\ref{tab:nist90b} presents the SP~800-90B results for all six
configurations, with key trends summarized in
Fig.~\ref{fig:sp80090b}. The compression estimator is the binding
constraint across the entire gain sweep:
$\hmin^{(\mathrm{all})}=\hmin^{(\mathrm{compression})}$ for every
configuration (Fig.~\ref{fig:sp80090b}b). The compression estimator
detects exploitable patterns in the sequence of packed bytes, so its
consistent dominance points to residual serial correlation in the
decimated bitstream as the common cause.

\begin{table*}[t]
\caption{NIST SP 800-90B Results (Bitstring Path, $1.50\!\times\!10^{6}$ bytes, Before VN Debiasing)}
\label{tab:nist90b}
\centering\small 
\begin{tabular}{@{}lcrrrrrrl@{}}
\toprule
Config & IID & $h_{\mathrm{comp}}^{\dagger}$ & $h_{\mathrm{mcv}}$ & $h_{\mathrm{markov}}$ & $h_{\mathrm{lrs}}$ & $\boldsymbol{h_{\mathrm{min}}}$ & Throughput & SP 800-90B \\
       &     & (b/b)      & (b/b)              & (b/b)                 & (b/b)              & \textbf{(b/b)}       & (b/s)      & Verdict \\
\midrule
$G\!=\!101$, $0.1$\textmu{}F  & \checkmark & 0.832 & 0.941 & 0.949 & 0.900 & \textbf{0.832} & 9{,}856  & Marginal \\
$G\!=\!48$, $0.1$\textmu{}F   & \checkmark & 0.801 & 0.929 & 0.935 & 0.995 & \textbf{0.801} & 11{,}568 & Marginal$^{\star}$ \\
$G\!=\!31$, $0.1$\textmu{}F   & \checkmark & 0.752 & 0.898 & 0.905 & 0.982 & \textbf{0.752} & 10{,}896 & Marginal \\
$G\!=\!23$, $0.1$\textmu{}F   & \checkmark & 0.675 & 0.850 & 0.857 & 0.982 & \textbf{0.675} & 9{,}537  & Insufficient \\
$G\!=\!11$, $0.1$\textmu{}F   & $\times$   & 0.441 & 0.650 & 0.654 & 0.895 & \textbf{0.441} & 7{,}640  & Insufficient \\
$G\!=\!101$, $0.01$\textmu{}F & \checkmark & 0.869 & 0.957 & 0.958 & 0.981 & \textbf{0.869} & 13{,}534 & Marginal \\
\bottomrule
\multicolumn{9}{@{}l}{\footnotesize $^{\dagger}$Compression estimator is binding ($h_{\mathrm{min}}^{(\mathrm{all})}=h_{\mathrm{compression}}$) for all six configurations.}\\
\multicolumn{9}{@{}l}{\footnotesize $^{\star}$Optimal: 15/15 SP 800-22 and peak Cg=0.1\textmu{}F throughput.}
\end{tabular}
\end{table*}
The gap between $h_{\mathrm{compression}}$ and the next estimator
($h_{\mathrm{MCV}}\!\approx\! 0.85\text{--}0.96$~b/b across
configurations) widens monotonically as gain decreases: 0.109~b/b at
$G=101$, rising to 0.209~b/b at $G=11$. This progressive widening reflects the increasing NTO rate at lower gain. The near-threshold oscillation produces correlated bit bursts that the compression estimator identifies as exploitable structure, while the distribution-based MCV estimator is insensitive to the temporal ordering of symbols. The gap therefore quantifies the entropic cost of near-threshold oscillation in a directly measurable form.

\subsection{IID Assessment}

IID failure is isolated exclusively to $G=11$. All five other
configurations pass the SP~800-90B permutation test, establishing a
binary IID boundary between $G=11$ and $G=23$. At $G=11$, the NTO rate of 29.2\% creates correlated bit
bursts sufficiently long and frequent to violate the IID assumption
at the permutation test's significance level. At $G=23$, the NTO
rate of 22.9\% remains below this threshold, suggesting that the IID
boundary lies in the NTO-rate range 22.9--29.2\%.

\subsection{Throughput Analysis}

Entropy throughput $\Phi=\favg\!\times\!\hmin$ is non-monotonic in
gain, peaking at $G=48$ (11{,}568~b/s) within the
$\Cg=\SI{0.1}{\micro\farad}$ sweep (Fig.~\ref{fig:sp80090b}c). $G=101$ delivers higher $\hmin$
(0.832~b/b) but lower $\favg$ (\SI{11841}{\hertz}), yielding lower
throughput (9{,}856~b/s). This non-monotonicity arises because the
switching frequency which sets the raw bit-generation rate before
decimation does not scale monotonically with gain. $\favg$ peaks
near $G=31$--$48$ (\SI{14449}{} -- \SI{14490}{\hertz}) before
declining at higher gain due to the MCP6002 GBW limitation and the
longer RC time constant at larger $\Rf$. The product
$\favg\!\times\!\hmin$ therefore peaks at $G=48$, where neither
factor is at its individual extreme but their product is maximized.
The circuit achieves overall maximum throughput of 13{,}534~b/s at $G=101$
with $\Cg=\SI{0.01}{\micro\farad}$, as analyzed in
Section~\ref{sec:cg}.

\section{Minimum Viable Gain Boundary}\label{sec:boundary}
This section establishes the minimum viable gain boundary from the
convergent SP~800-22 and SP~800-90B results and provides
gain-selection guidelines for deployment.
\subsection{Cross-Standard Convergence}

A distinguishing feature of this dataset is the independent
convergence of two NIST standards on consistent gain-quality
boundaries through entirely different mathematical mechanisms.
SP~800-90B's compression estimator detects temporal structure in
byte sequences. SP~800-22's T03 (Runs) test detects run-length
imbalance in the bit sequence, and the SP~800-90B permutation test
detects sequential structure absent in IID samples. All three respond to the same physical phenomenon; near-threshold oscillation producing correlated bit bursts. The agreement of
their findings is therefore not circular and provides stronger
evidence for the gain-boundary conclusions than either standard alone
could furnish.

\subsection{Gain Selection Guidelines}

Based on the dual-standard assessment, the following guidelines apply
to deployments using this circuit topology:

\textbf{Minimum viable ($G=23$, $\Rf=\SI{22}{\kilo\ohm}$):}
$\hmin=0.675$~b/b, IID pass, SP~800-90B Insufficient. Acceptable for
applications requiring entropy diversity without formal SP~800-90B
certification, where post-processing conditioning will be applied.

\textbf{Recommended minimum ($G=31$, $\Rf=\SI{30}{\kilo\ohm}$):}
$\hmin=0.752$~b/b, IID pass, SP~800-90B Marginal. The lowest gain
meeting the Marginal threshold; suitable for conditioned entropy
delivery.

\textbf{Optimal ($G=48$, $\Rf=\SI{47}{\kilo\ohm}$):} $\hmin=0.801$~b/b,
IID pass, 15/15 SP~800-22, peak throughput 11{,}568~b/s. Recommended
operating point balancing entropy quality, throughput, and standards
compliance.

\textbf{High entropy ($G=101$, $\Rf=\SI{100}{\kilo\ohm}$,
$\Cg=\SI{0.1}{\micro\farad}$):} $\hmin=0.832$~b/b, IID pass, 14/15
SP~800-22. Preferred when maximum $\hmin$ is prioritized over
throughput.

\textbf{Maximum throughput ($G=101$, $\Cg=\SI{0.01}{\micro\farad}$):}
$\hmin=0.869$~b/b, IID pass, 13/13 SP~800-22, 13{,}534~b/s.
Recommended when throughput is the primary constraint and a single
passive-component substitution is feasible.

The boundary $G=11$ is unambiguously inadequate by both standards:
IID failure, $\hmin=0.441$~b/b (SP~800-90B Insufficient), and the
highest NTO rate (29.2\%) of any tested configuration. $G=11$
should be excluded from any deployment requiring statistical
independence of the entropy-source output.

\section{Applicability to Industrial Control System Deployments}\label{sec:ics}
This section evaluates the applicability of the characterized TRNG to
industrial control system deployments, addressing entropy budgets,
conditioning, and integration cost.
\subsection{ICS Entropy Requirements}

Industrial control systems are increasingly subject to mandatory
cybersecurity frameworks that presuppose access to hardware entropy
sources at the field-device level.
IEC~62443-3-3~\cite{iec62443} Security Level~2 requires cryptographic
authentication for all inter-zone communications, which in practice
demands fresh nonces and session-specific key material at each
authentication event. NERC~CIP-015-1~\cite{nerc2024}, the emerging
internal-network security standard for bulk electric systems,
mandates random session identifiers for all authenticated remote
access sessions. IEC~62351-8~\cite{iec62351}, covering role-based
access control for power systems, requires entropy for privilege-token
generation. IEEE~1815 (DNP3)~\cite{ieee1815}, the dominant SCADA
communication protocol in North American electric utilities, mandates
random challenge values for Secure Authentication v5. Each of these
standards implicitly requires a hardware entropy source at the PLC or
IED level

The field-device entropy budget is constrained by scan-cycle timing.
ICS scan cycles range from \SI{1}{\milli\second} (substation
protection IEDs) to \SI{500}{\milli\second} (water-treatment SCADA
RTUs). At $G=48$, the TRNG delivers 11{,}568~b/s, providing 11.6 bits
per \SI{1}{\milli\second} cycle and 5{,}784 bits per
\SI{500}{\milli\second} cycle. A 128-bit session nonce accumulates in
\SI{11.1}{\milli\second}; a 256-bit key seed in
\SI{22.1}{\milli\second}. These accumulation times fit comfortably
within the timing budget of all ICS scan classes except the most
latency-critical protection-relaying applications (sub-\SI{1}{\milli%
\second}), which are typically handled by dedicated hardware security
modules in any case.

\subsection{Conditioning for Cryptographic Use}

The SP~800-90B Marginal verdict at $G=48$ reflects that the raw
source output requires conditioning before cryptographic-strength key
material can be derived. SP~800-90B Section~10 specifies approved
conditioning constructions; the simplest appropriate for this source
is the Hash\_df construction using SHA-256 \cite{nist80090a}. With $\hmin=0.801$~b/b
and 512 bits of TRNG input, Hash\_df delivers 256 bits of output with
full 256-bit security strength. At 11{,}568~b/s raw throughput, the
conditioner can be refreshed approximately 22 times per
second---more than adequate for all ICS authentication use cases. For
non-cryptographic applications (nonce diversity, challenge
randomization, seed diversification) the unconditioned TRNG output at
$G=48$ is directly usable.

\subsection{Integration Pathway and Cost}

The complete TRNG circuit (1N4749A, MT3608 module, MCP6002, passives)
has a bill-of-materials cost below \$5 at single-unit
pricing, representing a cost reduction of two to four orders of
magnitude relative to commercial HSMs. The circuit requires no
specialized PCB, shielding, or calibration; the gain-selection
guideline ($\Rf=\SI{47}{\kilo\ohm}$ for $G=48$) is a single resistor
specification. The digital output at Node~E is a 3.3--5~V
logic-level square wave compatible with GPIO inputs on all major PLC
families (Siemens S7-1200, Allen-Bradley MicroLogix, Schneider
Modicon, Click C0-series, OpenPLC) without level shifting.

Future work can integrate the entropy source with multi-vendor PLC testbeds. For each vendor, potential work can 
characterize the entropy-accumulation latency within IEC~61131-3
scan-cycle constraints and Modbus/TCP authenticated session-key
generation under operational traffic load. The present paper
establishes the entropy-source characterization that such integration
work requires as its foundation. Without knowing which gain setting
produces IID-compliant output at sufficient entropy rate, integration
experiments cannot be designed with statistical rigor. The
recommendation of $G=48$ as the optimal operating point provides that
foundation.

\section{Conclusion}\label{sec:conclusion}

This paper presents a dual-standard entropy characterization of a
Zener avalanche-noise TRNG across a gain sweep ($G=11$ to $G=101$)
and a coupling-capacitor study, using $1.44\!\times\!10^{9}$ raw bits
per configuration---the largest dataset reported for this circuit
family. A foundational question in any such characterization is
whether the observed switching arises from the intended noise source
or from amplifier self-excitation. A DC substitution experiment,
replacing the Zener diode with a stable \SI{23.5}{\volt} source at the
same Node~A bias point, resolved this directly: mean switching
frequency fell by approximately 97\%
($\SI{11841}{\hertz}\to\SI{407}{\hertz}$), establishing Zener
avalanche noise as the necessary entropy source across all
configurations. The coupling capacitor $\Cg$ does not alter this
attribution; it determines instead whether the Zener noise produces
near-threshold oscillation ($\Cg=\SI{0.1}{\micro\farad}$) or
committed threshold crossings ($\Cg=\SI{0.01}{\micro\farad}$). Both
regimes share a three-phase autocorrelation structure (positive,
anti-correlated, independent) attributed to MCP6002 saturation-recovery
dynamics, which motivates the $D=120$ decimation factor and places all
assessed samples in the statistically independent regime. Under
SP~800-90B, the compression estimator is the universally binding
min-entropy constraint, and its gap from $h_{\mathrm{MCV}}$-widening
from 0.109~b/b at $G=101$ to 0.209~b/b at $G=11$-directly
quantifies the entropic cost of near-threshold oscillation as gain
decreases. Reducing $\Cg$ to $\SI{0.01}{\micro\farad}$ at $G=101$
exploits committed-crossing dynamics to achieve a Pareto improvement
from a single passive-component change: $\hmin$ rises from 0.832 to
0.869~b/b, throughput from 9{,}856 to 13{,}534~b/s, and the
near-threshold oscillation rate falls from 13.3\% to 7.9\%.
Within the $\Cg=\SI{0.1}{\micro\farad}$ sweep, $G=48$ emerges as the
optimal operating point by the convergent evidence of both standards:
the sole 15/15 SP~800-22 score, $\hmin=0.801$~b/b, IID pass, and
peak throughput of 11{,}568~b/s at a BOM cost below~\$5. At the other
extreme, $G=11$ fails IID outright, placing the minimum viable gain
boundary at $G=23$.

Future work will characterize temperature dependence across the
industrial operating range ($-40\,\degC$ to $+85\,\degC$), evaluate
NodeREF trimmer integration for bias correction, extend the assessment
to the full SP~800-90B ten-estimator suite and temporal stationarity
under extended operation, and integrate the TRNG with multi-vendor PLC
testbeds---Siemens S7-1200, Allen-Bradley MicroLogix~1400, Schneider
Modicon~M221, and OpenPLC on Raspberry~Pi---under IEC~61131-3
scan-cycle constraints.


\end{document}